\documentclass{article}
\usepackage{ijcai19}

\usepackage{times}
\usepackage{soul}
\usepackage{url}
\usepackage[hidelinks]{hyperref}
\usepackage[utf8]{inputenc}
\usepackage[small]{caption}
\usepackage{graphicx}
\usepackage{booktabs}
\usepackage{color}
\usepackage{tikz}
\usepackage{float}
\usepackage{bbm}
\usepackage{caption}
\usepackage{subcaption}
\usepackage{pgfplots}
\usepackage{float}
\usepackage{amsmath,amssymb}
\usepackage{color}
\usepackage{booktabs}
\usepackage{multirow}
\usepackage{bbm}
\usepackage[linesnumbered,ruled]{algorithm2e}

\graphicspath{{./images/}{./photos/}{./logo/}}

\begin{document}

\title{Causal Embeddings for Recommendation: An Extended Abstract}

\author{
Stephen Bonner$^{1,2}$
\and
Flavian Vasile$^2$
\affiliations
$^1$Durham University, UK\\
$^2$Criteo AI Lab, Paris
\emails
s.a.r.bonner@durham.ac.uk,
f.vasile@criteo.com
}

\maketitle

\begin{abstract}
Recommendations are commonly used to modify user's natural behavior, for example, increasing product sales or the time spent on a website. This results in a gap between the ultimate business objective and the classical setup where recommendations are optimized to be coherent with past user behavior. To bridge this gap, we propose a new learning setup for recommendation that optimizes for the Incremental Treatment Effect (ITE) of the policy. We show this is equivalent to learning to predict recommendation outcomes under a fully random recommendation policy. We propose a new domain adaptation algorithm that learns from logged data containing outcomes from a biased recommendation policy and predicts recommendation outcomes according to behaviour under random exposure. We compare our method against state-of-the-art factorization methods, in addition to new approaches of causal recommendation and show significant improvements.
\end{abstract}

\section{Introduction}
\label{sec:intro}
In recent years, online commerce has outpaced the growth of traditional commerce. As such, research work on \emph{recommender systems} has also grown significantly, with recent \emph{Deep Learning (DL)} approaches achieving state-of-the-art results. Broadly, these DL approaches  frame the recommendation task as either:
  \begin{itemize}
    \item A distance learning problem between pairs of products or pairs of users and products, measured with Mean Squared Error (MSE) and Area Under the Curve (AUC), like in the work by \cite{prod2vec,metaprod2vec,glove}.
    \item A next item prediction problem that models user behavior and predicts the next action, measured with ranking metrics such as Precision@K and Normalized Discounted Cumulative Gain (NDCG), as presented in \cite{hidasi2015session,hidasi2016parallel}.
 \end{itemize}
    
However, we argue that both approaches fail to model the inherent interventionist nature of recommendation, which should not only attempt to model the organic user behavior, but to actually attempt to optimally influence it according to a preset objective.
    
Ideally, the change in user behavior should be measured against a case where no recommendations are shown. This is not an easy problem, since we do not know what the user would have done in the absence of recommendations and is a natural fit for the causal / counterfactual inference paradigm. 

Using a causal vocabulary, we are interested in finding the treatment recommendation policy that maximizes the reward obtained from each user with respect to the control recommendation policy. This objective is traditionally denoted as the Individual Treatment Effect (ITE) \cite{rubin1974estimating}. 


In our work, we introduce a modification to the classical matrix factorization approach which leverages both a large biased sample of biased recommendation outcomes and a small sample of randomized recommendation outcomes in order to create user and products representations. We show that using our method, the associated pairwise distance between user and item pairs is a more strongly aligned with the corresponding ITE of recommending a particular item to the user than in both traditional matrix factorization and causal inference approaches.
\subsection{Causal Vocabulary}
Below we briefly introduce the causal vocabulary and notation that we will be using throughout the paper.

\paragraph{The Causal Inference Objective.}

In the \emph{classical setup}, we want to determine the causal effect of one single action which constitutes the treatment versus the control case where no action or a placebo action is undertaken (do vs. not do). In the \emph{stochastic setup}, we want to determine the causal effect of a stochastic treatment policy versus the baseline control policy. In this case, both treatment and control are distributions over all possible actions. We retrieve the classical setup as a special case.
    
\paragraph{Recommendation Policy.}  We assume a \emph{stochastic policy} $\pi_x$ that associates to each user $u_i$ and product $p_j$ a probability for the user $u_i$ to be exposed to the recommendation of product $p_j$:
      $$p_j \sim \pi_{x}(.|u_i)$$
    
 For simplicity we assume showing no products is also a valid intervention in $\mathcal{P}$.

\paragraph{Policy Rewards.} Reward $r_{ij}$ is distributed according to an unknown conditional distribution $r$ depending on $u_i$ and $p_j$:
    $$
    r_{ij} \sim r(.|u_i,p_j)
    $$

The reward $R^{\pi_x}$ associated with a policy $\pi_x$ is equal to the sum of the rewards collected across all incoming users by using the associated personalized product exposure probability:
    $$
    R^{\pi_x} =  \sum_{ij}  r_{ij} \pi_x(p_j | u_i) p(u_i) = \sum_{ij} R_{ij}^{\pi_x}
    $$

\paragraph{Individual Treatment Effect.} The \emph{Individual Treatment Effect (ITE)} value of a policy $\pi_x$ for a given user $i$ and a product $j$ is defined as the difference between its reward and the control policy reward:
    $$ITE_{ij}^{\pi_x} =  R_{ij}^{\pi_x} - R_{ij}^{\pi_c}$$
   
  We are interested in finding the policy $\pi^{*}$ with the \emph{highest sum of ITEs}:
    $$
    \pi^{*} = arg\max_{\pi_x} \{ITE^{\pi_x} \}
    $$
    where: $ITE^{\pi_x} = \sum_{ij} ITE_{ij}^{\pi_x}$


\paragraph{Optimal ITE Policy.} It is easy to show that, starting from any control policy $\pi_c$, the best incremental policy $\pi^*$ is the policy that shows deterministically to each user $u_i$ the product $p_i^{*}$ with the highest personalized reward $r_i^{*}$:
  $$
   \pi^* =  \pi_{det}=
    \begin{cases}
        1,& \text{if } p_j = p_i^{*} \\
        0,              & \text{otherwise}
   \end{cases}
   $$
   
 Note: This assumes non-fatigability, e.g. non-diminishing returns of recommending the same product repeatedly to the user (no user state / repeated action effects at play).


\paragraph{IPS Solution For $\pi^*$}

 In order to find the optimal policy $\pi^*$ we need to find for each user $u_i$ the product with the highest personalized reward $r_i^{*}$.

In practice we do not observe directly  $r_{ij}$, but $y_{ij} \sim r_{ij} \pi_c(p_j | u_i)$.

The current approach to estimate $r_{ij}$ constitutes in using \emph{Inverse Propensity Scoring (IPS)}-based methods to predict the unobserved reward $r_{ij}$:

    $$\hat{r}_{ij} \approx \frac{y_{ij}}{\pi_c(p_j | u_i)}$$
    
This assumes we have incorporated randomization in the current policy $\pi_c$. Even with the existence of randomization, the main shortcoming of IPS-based estimators is that they do not handle well big shifts in exposure probability between treatment and control policies (products with low probability under the logging policy $\pi_c$ will tend to have higher predicted rewards).
 
\paragraph{Addressing the variance issues Of IPS.} It is easy to observe that in order to obtain minimum variance we should collect data using fully randomized recommendations, e.g. when: $\pi_c = \pi_{rand}$. However, this means zero recommendation performance and therefore cannot be a solution in practice.
    
\textbf{Our question:} Could we learn from $\pi_c$ a predictor for performance under $\pi_{rand}$ and use it to compute the optimal product recommendations $p_i^{*}$?
\section{Our Approach: Causal Embeddings (CausE)}

 We are interested in building a good predictor for recommendation outcomes under random exposure for all the user-product pairs, which we denote as $\hat{y}_{ij}^{rand}$. We make the assumption that we have access to a large sample $S_c$ from the logging policy $\pi_c$ and a small sample $S_t$ from the randomized treatment policy $\pi_{t=rand}$ (e.g. the logging policy $\pi_c$ uses \emph{e-greedy} randomization).
  
 To this end, we propose a multi-task objective that jointly factorizes the matrix of observations $y^c_{ij} \in S_c$ and the matrix of observations $y^t_{ij} \in S_t$. Our approach is inspired by the work in \cite{rosenfeld2016predicting} and shares similarities with other domain-adaptation based models for counterfactual inference such as the work in \cite{johansson2016learning,shalit2017estimating}. 
		
\subsection{Predicting Rewards Via Matrix Factorization}

\begin{figure}[h!]
			\includegraphics[scale=0.22]{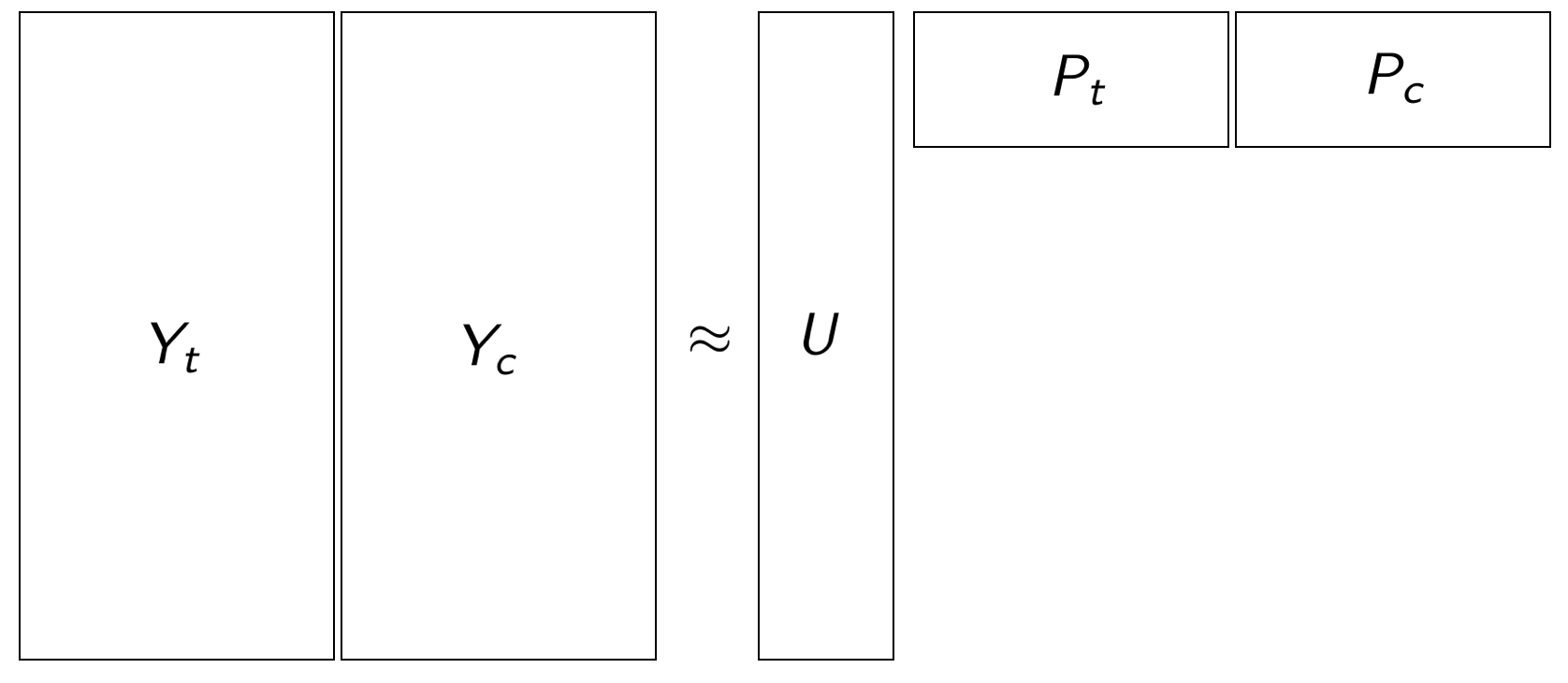}
    		\centering
    		\caption{The joint MF problem.}
    		\label{fig:joint_mf}
    		\vskip -10pt
		\end{figure}

By using a matrix factorization model, we assume that both the expected factual control and treatment rewards can be approximated as linear predictors over the \textbf{shared} user representations $u_i$, as shown in Fig. \ref{fig:joint_mf}.
    $$
    y^{c}_{ij} \approx <p^{c}_j,u_i>
    $$
    $$
    y^{t}_{ij} \approx <p^{t}_j,u_i>
    $$

   As a result, we can approximate the ITE of a user-product pair $i,j$ as the difference between the two, see eq.\ref{eq:ite_mf} below:
    
    \begin{equation}
    \widehat{ITE}_{ij} = <p^t_j,u_i> - <p^c_j,u_i> = <w^{\Delta}_j, u_i>  
    \label{eq:ite_mf}
    \end{equation}

 \paragraph{Proposed joint optimization solution}
 
 The joint optimization objective has naturally two terms, one measuring the performance of the solution on the treatment sample and the on control sample. The novel part of the objective comes from the additional constraint on the distance between the treatment and control vectors for the same action/item, that can be directly linked to the ITE effect of the item. We are listing below each one of the terms of the overall objective. 
 
  \paragraph{Sub-objective \#1: Treatment Loss Term $L_t$}

  We define the first part of our joint prediction objective as the supervised predictor for $y^t_{ij}$, trained on the limited sample $S_t$, as shown in the eq. \ref{eq:lt_mf} below:
    
    \begin{equation}
    L_{t}(P_t) = \sum_{(i,j,y_{ij}) \in S_t} l^t_{ij} = L(U P_t  , Y_t) + \Omega(P_t)
    \label{eq:lt_mf}
    \end{equation}
    
    where: 
    \begin{itemize}
    \item $P_t$ is the parameter matrix of treatment product representations.
    \item $U$ is the fixed matrix of the user representations. 
    \item $Y_t$ is the observed rewards matrix.  
    \item $L$ is an arbitrary loss function. 
    \item $\Omega(.)$ is a regularization term over the model parameters. 

  \end{itemize}


  \paragraph{Linking the control and treatment effects}

 Additionally, we can use the translation factor in order to be able to use the model built from the treatment data $S_t$ to predict outcomes from the control distribution $S_c$: $$<p^c_j,u_i> = <p^t_j - w^{\Delta}_j,u_i>$$

  \paragraph{Sub-objective \#2: Control Loss Term $L_c$ }

 Now we want to leverage our ample control data $S_c$ and we can use our treatment product representations through a translation:
    \begin{equation*}
    \begin{split}
    L_{c}(P_t, W^\Delta) = \sum_{(i,j,y_{ij}) \in S_c} l^c_{ij}  \\
    = L(U (P_t -  W^\Delta), Y_c) + \Omega(P_t , W^\Delta)
    \end{split}
    \end{equation*}
    
    which can be written equivalently as:  
    \begin{equation}
    L_{c}(P_t,P_c) = \sum_{(i,j,y_{ij}) \in S_c} l^c_{ij} = L(U P_c  , Y_c) + \Omega(P_c , W^\Delta)
    \label{eq:lc_mf}
    \end{equation}
    
 where we regularize the control $P_c$ against the treatment embeddings $P_t$ ($W^\Delta = P_t - P_c$). As shown in the eq. \ref{eq:ips_wdelta} below, we can see that $IPS$ is a function of $W^{\Delta}$. Therefore, by regularizing $W^{\Delta}$ we are effectively putting a constraint on the magnitude of the $IPS$ term.

  \begin{equation}
  IPS_{ij} = \frac{\pi_t(p_j| u_i)}{\pi_c(p_j| u_i)} = \frac{<u_i, p^{t}_j>}{<u_i, p^{c}_j>} = 1 + \frac{<u_i, w^{\Delta}_j>}{<u_i, p^{c}_j>} 
  \label{eq:ips_wdelta}
  \end{equation}
  

\paragraph{Overall Joint Objective} By putting the two tasks together ($L_t$ and $L_c$) and regrouping the loss functions and the regularizer terms, we have that:
    \begin{equation}
    \begin{split}
        L^{prod}_{CausE}(P_t, P_c)= L(P_t, P_c) \\  + \Omega_{disc}(P_t - P_c) + \Omega_{embed}(P_t,P_c)
    \end{split}
    \end{equation}
where $L(.)$ is the reconstruction loss function for the concatenation matrix of $P_t$ and $P_c$, $\Omega_{disc}(.)$ is a regularization function that weights the discrepancy between the treatment and control product representations and $\Omega_{embed}(.)$ is a regularization function that weights the representation vectors.

		\begin{figure}[h!]
			\includegraphics[scale=0.22]{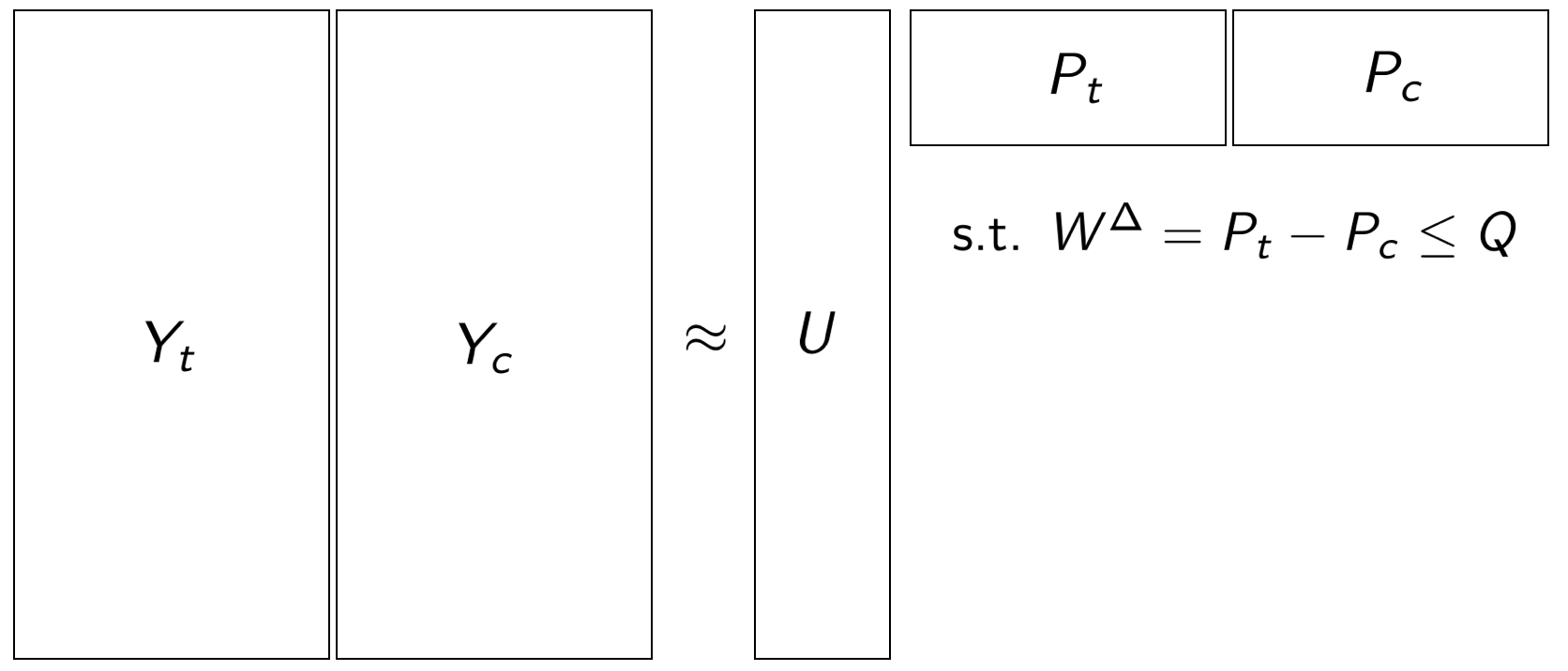}
    		\centering
    		\caption{The final joint MF objective.}
    		\vskip -10pt
    		\label{fig:full_mf}
		\end{figure}

\paragraph{Question: How about user shift?} The current recommendation solution is targeting a subset of users, for example, active buyers on a website and the new recommendation targets mainly newly signed users (modulo randomization which should give non-zero probabilities for all user product pairs).

\paragraph{Generalization of the objective to user shift} Our objective function can be altered to allow for the user representations to change, we obtain the equation below:
 
\begin{equation*}
    \begin{split}
    L^{user}_{CausE}(U_t, U_c) = L(U_t, U_c) \\
    + \Omega_{disc}(U_t - U_c) + \Omega_{embed}(U_t,U_c)
        \end{split}
\end{equation*}

Putting the loss functions associated with the user and product dimension together ($L^{prod}_{CausE}$,  $L^{user}_{CausE}$), we reach the final loss function for our method:
    \begin{equation}
    \begin{split}
    L_{CausE}(P_t, P_c, U_t, U_c) = L(P_t, P_c, U_t, U_c) \\
    +\Omega_{disc}(P_t - P_c, U_t - Uc) + \Omega_{embed}(P_t, P_c, U_t,U_c)
    \end{split}
    \end{equation}


\footnotesize

 \begin{algorithm}[h]
    \SetKwInOut{Input}{Input}
    \SetKwInOut{Output}{Output}
    
    \Input{Mini-batches of $S_c = { \{ (u_i, p^c_j, \delta^c_{ij}) \} }^{M_c}_{i=1}$ and $S_t = { \{ (u_i, p^t_j, \delta^t_{ij}) \} }^{M_t}_{i=1}$ , regularization parameters $\lambda_{embed}$ and $\lambda_{dist}$, learning rate $\eta$}
    
    \Output{$P_t, P_c, U_t, U_c$  - Product and User Control and Treatment Matrices} 
    Random initialization of $P_t, P_c, U_t, U_c$ \;
    \While{not converged}{
      read batch of training samples\;
          \For{\textbf{each} product $p_j$ in $P_c, P_t$}{
        Update product vector: $p_j \leftarrow p_j - \eta \nabla L^{prod}_{CausE}(p,\lambda_{embed},\lambda_{dist})$)
        }
      \For{\textbf{each} user $u_i$ in $U_c, U_t$}{
          Update user vector: $u_i \leftarrow u_i - \eta \nabla L^{user}_{CausE}(u,\lambda_{embed},\lambda_{dist})$)
        }
      }
      \Return{$P_t, P_c, U_t, U_c$}
    \caption{CausE Algorithm: Causal Embeddings For Recommendations}
    \label{algo:cause}
  \end{algorithm}
\section{Experimental Results}

\subsection{Experimental Setup}
The task is predicting the outcomes $y^{rand}_{ij}$ under treatment policy $\pi_{rand}$, where all of the methods have available at training time a large sample of observed recommendations outcomes from $\pi_c$ and a small sample from $\pi_{rand}$. Essentially this is a classical conversion-rate prediction problem so we measure \emph{Mean-Squared Error (MSE)} and \emph{Negative Log-Likelihood (NLL)}. We report lift over average conversation rate from the test dataset: 
  $$
    \label{eq:lift}
   lift_x^{metric} =  \frac{metric_x - metric_{AvgCR}}{metric_{AvgCR}} 
  $$
\begin{table*}[t!]
  \centering
  \resizebox{\textwidth}{!}{%
  \begin{tabular}{@{}l c c c c c c @{}}
  \toprule
  \textbf{Method}      & \multicolumn{3}{c}{\textbf{MovieLens10M (SKEW)}}  & \multicolumn{3}{c}{\textbf{Netflix (SKEW)}}    \\ \midrule
              & \textbf{MSE lift}  & \textbf{NLL lift} & \textbf{AUC}   & \textbf{MSE lift}     & \textbf{NLL lift}   & \textbf{AUC} \\
  
  \textit{BPR-no}  & $-$  & $-$ & $0.693 (\pm 0.001)$ & $-$ & $-$ & $0.665 (\pm0.001)$   \\
  \textit{BPR-blend}  & $-$  & $-$ & $0.711 (\pm 0.001)$ & $-$ & $-$ & $0.671 (\pm0.001)$   \\
  \textit{SP2V-no}  & $+3.94\% (\pm 0.04)$  & $+4.50\% (\pm 0.04)$ & $0.757 (\pm 0.001)$ & $+10.82\% (\pm0.02)$ & $+10.19\% (\pm0.01)$ & $0.752(\pm0.002)$   \\
  \textit{SP2V-blend} & $+4.37\% (\pm 0.04)$  & $+5.01\% (\pm 0.05)$ & $0.768 (\pm 0.001)$ & $+12.82\% (\pm0.02)$ & $+11.54\% (\pm0.02)$ & $0.764(\pm0.003)$   \\
  \textit{SP2V-test}  & $+2.45\% (\pm 0.02)$  & $+3.56\% (\pm 0.02)$ & $0.741 (\pm 0.001)$ & $+05.67\% (\pm0.02)$ & $+06.23\% (\pm0.02)$ & $0.739(\pm0.004)$   \\
  \textit{WSP2V-no}   & $+5.66\% (\pm 0.03)$  & $+7.44\% (\pm 0.03)$ & $0.786 (\pm 0.001)$ & $+13.52\% (\pm0.01)$ & $+13.11\% (\pm0.01)$ & $0.779(\pm0.001)$   \\
  \textit{WSP2V-blend}  & $+6.14\% (\pm 0.03)$  & $+8.05\% (\pm 0.03)$ & $0.792 (\pm 0.001)$ & $+14.72\% (\pm0.02)$ & $+14.23\% (\pm0.02)$ & $0.782 (\pm0.002)$   \\
  \textit{BN-blend}  & $-$  & $-$ & $0.794 (\pm 0.001)$ & $-$ & $-$ & $0.785 (\pm0.001)$   \\
  \hline
  \hline
  \textit{CausE-avg}  & $+12.67\% (\pm 0.09)$  & $+15.15\% (\pm 0.08)$ & $0.804 (\pm0.001)$ & $+15.62\% (\pm0.02)$ & $+15.21\% (\pm0.02)$ & $0.799 (\pm0.002)$   \\
  \textit{CausE-prod-T}  & $+07.46\% (\pm 0.08)$  & $+10.44\% (\pm 0.09)$ & $0.779 (\pm0.001)$ & $+13.97\% (\pm0.02)$ & $+13.52\% (\pm0.02)$ & $0.789 (\pm0.003)$   \\
  \hline
  \textbf{CausE-prod-C}  & $\mathbf{+15.48\% (\pm 0.09)}$  & $\mathbf{+19.12\% (\pm 0.08)}$ & $\mathbf{0.814 (\pm0.001)}$ & $\mathbf{+17.82\% (\pm0.02)}$ & $\mathbf{+17.19\% (\pm0.02)}$ & $\mathbf{0.821 (\pm0.003)}$   \\
  \bottomrule
  \end{tabular}}
  \caption{Results for MovieLens10M and Netflix on the Skewed (SKEW) test datasets. All three versions of the \emph{CausE} algorithm outperform both the standard and the IPS-weighted causal factorization methods, with \textit{CausE-avg} and \textit{CausE-prod-C} also out-performing BanditNet. We can observe that our best approach \emph{CausE-prod-C} outperforms the best competing approaches \emph{WSP2V-blend} by a large margin (21\% MSE and 20\% NLL lifts on the MovieLens10M dataset) and \textit{BN-blend} (5\% AUC lift on MovieLens10M).}
  \label{tab:results}
  \vskip -15pt
  \end{table*}

\newpage
 \subsection{Baselines}
We compare our method with the following baselines:
\paragraph{Matrix Factorization Baselines:}
\begin{itemize}
\item \textbf{Bayesian Personalized Ranking (BPR)} To compare our approach against a ranking based method, we use Bayesian Personalized Ranking (BPR) for matrix factorization on implicit feedback data \cite{rendle2009bpr}.
\item \textbf{Supervised-Prod2Vec (SP2V):} As a second factorization baseline we will use a Factorization Machine-like method \cite{rendle2010factorization} that approximates $y_{ij}$ as a sigmoid over a linear transform of the inner-product between the user and product representations.
\end{itemize}

\paragraph{Causal Inference Baselines:}
\begin{itemize}
\item \textbf{Weighted-SupervisedP2V (WSP2V):} We employ the SP2V algorithm on propensity-weighted data, this method is similar to the Propensity-Scored Matrix Factorization (PMF) from \cite{schnabel2016recommendations} but with cross-entropy reconstruction loss instead of MSE/MAE.

\item \textbf{BanditNet (BN):} To utilize BanditNet \cite{joachims2-18deep} as a baseline, we use SP2V as our target policy $\pi_w$. For the existing policy $\pi_c$, we model the behavior of the recommendation system as a popularity-based solution, described by the marginal probability of each product in the training data.
\end{itemize}

\subsection{Experimental Datasets}

We use the \emph{Netflix} and \emph{MovieLens10M} explicit rating datasets (1-5). In order to validate our method, we preprocess them as follows: We binarize the ratings $y_{ij}$ by setting 5-star ratings to 1 (click) and everything else to zero (view only) and generate a skewed dataset (SKEW) with 70/10/20 train/validation/test event split that simulates rewards collected from uniform exposure $\pi_t^{rand}$, following a similar protocol with the one presented in previous counterfactual estimation work such as in \cite{liangcausal,swaminathan2015batch} and described in detail in the long version of our paper \cite{bonner2018causal}. 


 \subsubsection{Experimental Setup: Exploration Sample $S_t$}
  We define 5 possible setups of incorporating the exploration data: 
  \begin{itemize}
    \item \textbf{No adaptation} \textit{(no)} - trained only on $S_c$.
    \item \textbf{Blended adaptation} \textit{(blend)} - trained on the blend of the $S_c$ and $S_t$ samples.
    \item \textbf{Test adaptation} \textit{(test)} - trained only on the $S_t$ samples.
    \item \textbf{Product adaptation} \textit{(prod)} - separate treatment embedding for each product based on the $S_t$ sample.
    \item \textbf{Average adaptation} \textit{(avg)} - average treatment product by pooling all the $S_t$ sample into a single vector.
\end{itemize}

\subsection{Results}

\begin{figure}[h!]
    \includegraphics[scale=0.4]{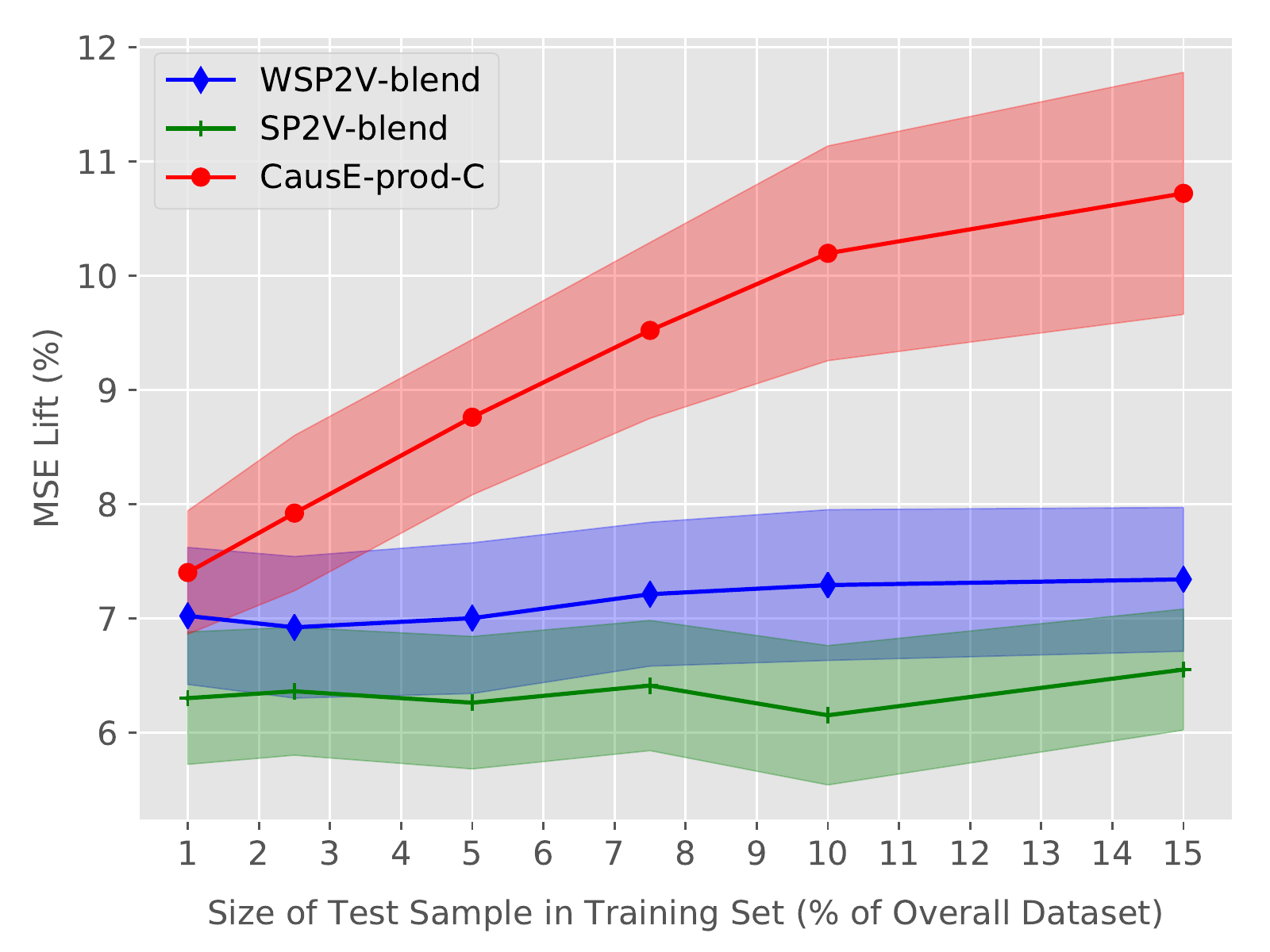}
      \centering
      \caption{Change in MSE lift as more test set is injected into the blend training dataset.}
      \vskip -15pt
      \label{fig:control}
  \end{figure}

Table \ref{tab:results} displays the results for running all the approaches on the datasets. Our proposed \textit{CausE} method significantly outperforms all baselines across both datasets, demonstrating that it has a better capacity to leverage the small test distribution sample $S_t$. We observe that, out of the three \emph{CausE} variants,  \emph{CausE-prod-C}, the variant that is using the regularized control matrix, clearly out-performs the others. Further, figure \ref{fig:control} highlights how CausE is able to make better use of increasing quantities of test distribution present in the training data compared with the baselines. 

\section{Conclusions}
\label{sec:conclusion}

We have introduced a novel method for factorizing matrices of user implicit feedback that optimizes for causal recommendation outcomes. We show that the objective of optimizing for causal recommendations is equivalent with factorizing a matrix of user responses collected under uniform exposure to item recommendations. We propose the \emph{CausE} algorithm, which is a simple extension to current matrix factorization algorithms that adds a regularizer term on the discrepancy between the item vectors used to fit the biased sample $S_c$ and the vectors that fit the uniform exposure sample $S_t$.  


\bibliography{literature}
\bibliographystyle{named}

\maketitle

\end{document}